\documentclass[twocolumn,showpacs,aps,prl,superscriptaddress]{revtex4-1}
\usepackage{amsfonts}
\usepackage{amsmath}
\usepackage{amssymb}
\usepackage{mathrsfs}
\usepackage{extarrows}
\usepackage{bm}
\usepackage{graphicx}
\usepackage{overpic}
\usepackage{color}
\usepackage{csquotes}

\newcommand{\nl}{\nonumber \\}
\newcommand{\be}{\begin{equation}}
\newcommand{\ee}{\end{equation}}
\newcommand{\bea}{\begin{eqnarray}}
\newcommand{\eea}{\end{eqnarray}}
\newcommand{\Fig}[1]{Fig.~\ref{#1}}

\newcommand{\Eq}[1]{Eq.\,(\ref{#1})}
\newcommand{\Eqs}[1]{Eqs.\,(\ref{#1})}

\newcommand{\Sch}{Schr\"{o}dinger}
\newcommand{\la}{\langle}
\newcommand{\ra}{\rangle}

\newcommand{\ti}{\tilde}

\newcommand{\iChEM}{{\it i}{\rm ChEM}}%

\newcommand{\w}{\omega}
\newcommand{\ep}{\epsilon}

\newcommand{\hc}{\hat{c}}
\newcommand{\hd}{\hat{d}}
\newcommand{\hF}{\hat{F}}
\newcommand{\bpsi}{\bar{\psi}}

\newcommand{\bareta}{\bar{\eta}}

\newcommand{\rhos}{\rho_{_{\rm S}}}
\newcommand{\rhob}{\rho_{_{\rm B}}}
\newcommand{\rhot}{\rho_{_{\rm T}}}
\newcommand{\Hs}{H_{_{\rm S}}}
\newcommand{\Hb}{H_{_{\rm B}}}
\newcommand{\Hsb}{H_{_{\rm SB}}}
\newcommand{\Vs}{V_{_{\rm S}}}

\newcommand{\trhos}{\tilde{\rho}_{_{\rm S}}}

\newcommand{\trs}{{\rm tr}_{_{\rm S}}}
\newcommand{\trb}{{\rm tr}_{_{\rm B}}}
\newcommand{\gm}{g^{-}}
\newcommand{\gp}{g^{+}}
\newcommand{\Cm}{C^{-}}
\newcommand{\Cp}{C^{+}}

\newcommand{\Xu}{X^+}
\newcommand{\Xd}{X^-}
\newcommand{\ty}{\tilde{y}}

\newcommand{\tgm}{\tilde{g}^-}
\newcommand{\tgp}{\tilde{g}^+}
\newcommand{\trho}{\tilde{\rho}}

\newcommand{\mM}{\mathcal{M}}
\newcommand{\mD}{\mathcal{D}}
\newcommand{\mB}{\mathcal{B}}

\makeatletter

\begin{document}

\title{Stochastic Representation of Non-Markovian Fermionic Quantum Dissipation}

\author{Lu Han}
\affiliation{Hefei National Laboratory for Physical Sciences at the Microscale \&
Synergetic Innovation Center of Quantum Information and Quantum Physics \&
CAS Center for Excellence in Nanoscience,
University of Science and Technology of China, Hefei, Anhui 230026, China}

\author{Vladimir Chernyak}
\affiliation{Hefei National Laboratory for Physical Sciences at the Microscale \&
Synergetic Innovation Center of Quantum Information and Quantum Physics \&
CAS Center for Excellence in Nanoscience,
University of Science and Technology of China, Hefei, Anhui 230026, China}
\affiliation{Department of Chemistry, Wayne State University, 5101 Cass Avenue, Detroit, MI 48202}

\author{Yun-An Yan}
\affiliation{School of Physics and Optoelectronic Engineering, Ludong University, Shandong 264025, China}

\author{Xiao Zheng} \email{xz58@ustc.edu.cn}
\affiliation{Hefei National Laboratory for Physical Sciences at the Microscale \&
Synergetic Innovation Center of Quantum Information and Quantum Physics \&
CAS Center for Excellence in Nanoscience,
University of Science and Technology of China, Hefei, Anhui 230026, China}

\author{YiJing Yan}
\affiliation{Hefei National Laboratory for Physical Sciences at the
Microscale \& \iChEM, University of Science and Technology of China, Hefei, Anhui
230026, China}

\begin{abstract}
  Quantum Brownian motion plays a fundamental role in many areas of modern physics.
  In the path-integral formulation, the environmental quantum fluctuations driving
  the system dynamics can be characterized by auxiliary stochastic fields.
  For fermion bath environment the stochastic fields are Grassmann-valued,
  and cannot be represented by conventional classical numbers.
  In this Letter, we propose a strategy to map the nonclassical Grassmann fields
  onto Gaussian white noises along with a set of quantized pseudo-states.
  This results in a numerically feasible stochastic equation of motion (SEOM) method
  for fermionic open systems.
  The SEOM yields exact physical observables for noninteracting systems,
  and accurate approximate results for interacting systems.
  The practicality and accuracy of the proposed SEOM are exemplified by direct
  stochastic simulations conducted on a single-impurity Anderson model.
\end{abstract}

\date{\today}


\maketitle

Over a century ago, Einstein has explained the nature of Brownian motion
by establishing a quantitative relation between the dissipative forces
driving a classical particle and the environmental thermal fluctuations \cite{Ein05549}.
Nowadays, quantum Brownian motion \cite{Cal83587,Ris85471,Gar88115,Kob0358,Han05026101},
\emph{i.e.}, the dissipative dynamics
of a quantum system driven by quantum fluctuations in surrounding environments,
plays a fundamental role in many subdisciplines of modern physics.
The sources of quantum fluctuations are rather general, \emph{e.g.},
the excitations of various types of particles or quasiparticles
(photons, phonons, electrons, excitons, spins, etc.)
This makes quantum Brownian motion closely pertinent to
a wide range of applications, including nanoelectronics, nanomotors,
solar energy conversion, superconductors, quantum information,
and quantum computation.

The main challenge in describing quantum Brownian motion
or more general quantum dissipative dynamics
is to elucidate the combined effects of
system-environment coupling, many-body correlation,
and non-Markovian memory \cite{Bre16021002,Veg17015001,Tam18030402}.
This requires a complete characterization of the influence of environment
which hosts the quantum fluctuations
and usually has infinite degrees of freedom.
Remarkable progress has been made, much thanks to the
path-integral formulation developed by Feynman and Vernon \cite{Fey63118}.
Particularly, it has been proposed that the influence of
environment can be captured by
introducing a set of auxiliary stochastic fields \cite{Kle95224}.

If the environment is a boson bath,
the auxiliary stochastic fields representing the environmental
quantum fluctuations are \emph{classical},
and can be realized via c-number noises.
Such formal simplicity has greatly facilitated the
development of stochastic theories.
For instance, quantum state diffusion (QSD) theory
\cite{Dio882885,Gis925677,Dio97569,Dio981699,Per99,Str991801,
Jin10240403,Sue14150403,Lin17180401}
and stochastic equation of motion (SEOM) theory
\cite{Sto982657,Sto02170407,Sha045053,Zho05334,Moi13134106,Zhu13095020,Yan18042126}
have been established and applied to investigate quantum
dissipative dynamics in realistic systems,
such as the transfer of excitons in molecular aggregates \cite{Jan09058301}
and photo-induced electron transfer at interfaces of
organic solar cells \cite{Abr1632914}.

While a boson mode corresponds to a classical harmonic oscillator,
there is no such classical counterpart for a fermion mode.
Therefore, the situation is highly nontrivial for fermion bath environment,
for which the auxiliary stochastic fields have to be Grassmann-valued, \emph{i.e.},
the fields mutually anticommute, to preserve the even parity of all physical
observables.
Unlike c-numbers, Grassmann numbers are \emph{nonclassical}
and cannot be represented by conventional means.
Such difficulty has severely hindered the practical
implementation of stochastic theories.
From the early attempts on describing fermion Brownian motion
\cite{Bar82172,App84473,Rog87353}
to the recent extension of QSD
\cite{Zha12032116,Che13052108,Shi13052127,Che14052104,Zha17121}
and SEOM \cite{Hsi18014103}
theories to fermionic open systems,
all previous efforts were limited to formal derivations \cite{Zha12032116,Hsi18014103};
whereas to the best of our knowledge, no stochastic simulation
has been conducted on quantum dissipative dynamics of fermions.

\begin{figure}[t]
\includegraphics[width=\columnwidth]{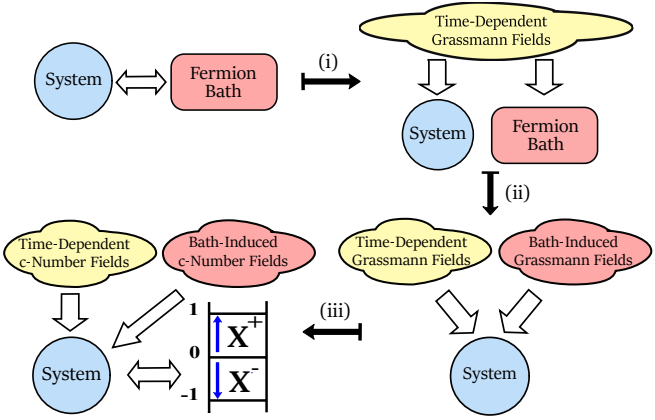}
\caption{Schematic illustration of the establishment of
  a stochastic method for fermionic open systems proposed in this Letter.
  After a series of mappings, labeled by (i), (ii) and (iii),
  the influence of fermion bath  on the system dynamics is finally captured by
  conventional c-number fields and a set of ladders with each
   consisting of three pseudo-states.}
\label{fig1}
\end{figure}

To break the status quo, and to enable direct numerical
simulation of fermion Brownian motion,
we propose in this Letter a mapping strategy,
as schematically illustrated in \Fig{fig1} and
the elaborations below,
with which the stochastic Grassmann fields are
effectively mapped onto conventional c-number fields
along with a set of quantized pseudo-states.
This finally leads to the construction of a numerically feasible
and accurate SEOM for fermionic open systems.

Without loss of generality, we illustrate
our strategy with a single-level system 
coupled linearly to a fermion bath. 
The system-bath interaction is
$\Hsb = \hc^\dag \hF + \hF^\dag \hc$,
with $\hF = \sum_k t_k \hd_k$.
Here, $\{t_k\}$ are coupling strengths;
$\hc$ ($\hc^\dag$) and $\hd_k$ ($\hd_k^\dag$)
are the annihilation (creation) operators
for system and bath levels, respectively.
Set $\hbar=1$ hereafter.
Following the standard stochastic decoupling method \cite{Sha045053},
we factorize $e^{-i\Hsb {\rm d}t}$
by introducing time-dependent auxiliary stochastic fields.
To fulfill the fermion statistics,
we introduce four involutive Grassmann--Wiener processes
 $\{\dot{\varphi}_{jt} = \eta_{jt}; j=1,\cdots,4\}$,
which satisfy $\la \eta_{jt} \ra = \la \bareta_{jt} \ra = 0$
and $\la \eta_{jt}\bareta_{j'\tau} \ra = \delta_{jj'}\delta(t-\tau)$.
The resultant decoupled system and bath density matrix dynamics
under the time-dependent auxiliary Grassmann fields (AGFs) read \cite{SM1}
\begin{align}
  \dot{\rho}_{_{\rm S}}
&= -i[\Hs, \rhos] + e^{-i\pi/4}
  \big(\hc^\dag \eta_{1t} + \bareta_{2t} \hc \big)\rhos
\nl&\quad\,
  + e^{i\pi/4} \rhos \big(\hc^\dag \eta_{3t} + \bareta_{4t} \hc \big),
\label{eom-rhos} \\
  \dot{\rho}_{_{\rm B}}
&= -i[\Hb, \rhob]
   + e^{-i\pi/4}\big(\bareta_{1t} \hF + \hF^\dag \eta_{2t} \big)\rhob
\nl&\quad\,
  + e^{i\pi/4} \rhob
  \big(\bareta_{3t} \hF + \hF^\dag \eta_{4t} \big).
  \label{eom-rhob}
\end{align}
Here, $\Hs$ ($\Hb$) is the system (bath) Hamiltonian.
The total density matrix of the system-and-bath composite can be
expressed as the stochastic average on the product of the solutions
to \Eqs{eom-rhos} and \eqref{eom-rhob}, \emph{i.e.},
\be \label{rho-tot}
\rhot = \la \rhos \rhob \ra
=\int_{t_0}^t \mathcal{D}\bm\bareta \,\mathcal{D}\bm\eta
 \, e^{-\int_{t_0}^t \bm{\bareta}_{\tau} \bm{\eta}_{\tau} {\rm d}\tau}
 \rhos \rhob.
\ee
Here, $\mathcal{D}\bm\bareta \,\mathcal{D}\bm\eta
\, e^{-\int_{t_0}^t \bm\bareta_{\tau} \bm\eta_{\tau} {\rm d}\tau}$
is the Grassmann--Wiener measure, with
$\bm\bareta\equiv \{\bareta_{j\tau}\}$ and $\bm\eta \equiv \{\eta_{j\tau}\}$.
By using the It\^{o}'s formula \cite{Klo92}, one can easily verify
that \Eqs{eom-rhos} and \eqref{eom-rhob},
the result of mapping (i) in \Fig{fig1},
recover exactly the \Sch\ equation for $\rhot$.

The primary goal of theoretical formulation
is to acquire the reduced system density matrix
$\rho = \trb(\rhot) = \la \rhos (\trb \rhob) \ra$,
with which the expectation value of any
system observable, $O = \trs(\hat{O} \rho)$,  can be evaluated.
%
Here, $\trb$ ($\trs$) denotes the trace over the bath (system)
subspace.
For a noninteracting fermion bath being initially in thermal equilibrium,
one can formally solve \Eq{eom-rhob} via the Magnus expansion \cite{Tan06},
and explicitly evaluate $\trb(\rhob)$ \cite{SM1}.
The influence of bath on reduced system dynamics
can be represented in a compact way, by introducing
bath-induced Grassmann fields, $g^\pm_t$,
which depend on the original AGFs in a linear yet
time-nonlocal manner:
%
%
\be\label{def-gt}
\begin{split}
  g^-_t & =  \int_{t_0}^t
 \left\{ [\Cp(t-\tau)]^\ast \eta_{4\tau}
 -i \Cm (t-\tau) \,\eta_{2\tau} \!  \right\} {\rm d}\tau,
\\
  g^+_t & = \int_{t_0}^t
 \left\{ [\Cm(t-\tau)]^\ast \bareta_{3\tau}
 -i \Cp (t-\tau) \,\bareta_{1\tau} \! \right\} {\rm d}\tau.
\end{split}
\ee
Here, $\Cp(t-\tau) = \trb \big[\hF^\dag(t)\, \hF(\tau)\, \rhob^{\rm eq}\big]$ and
$\Cm(t-\tau) = \trb \big[\hF(t)\, \hF^\dag(\tau)\, \rhob^{\rm eq}\big]$
are bath correlation functions, with
$\hF(t) \equiv e^{i\Hb t}\hF e^{-i\Hb t}$.
The reduced system density matrix is given by
$\rho = \la \trhos \ra$, with $\trhos$ satisfying
the following rigorous SEOM:
\begin{align}
 \dot{\tilde{\rho}}_{_{\rm S}}
&= -i[ \Hs, \trhos]
   + e^{-i\pi/4} \big\{\hc^\dag \gm_t - \gp_t \hc, \,\trhos \big\}
\nl & \quad
+ e^{-i\pi/4}
  \big(\hc^\dag \eta_{1t} + \bareta_{2t} \hc \big)\trhos
 + e^{i\pi/4} \trhos \big(\hc^\dag \eta_{3t} + \bareta_{4t} \hc \big) .
\label{eom-brhos}
\end{align}
This is the result of mapping (ii) in \Fig{fig1}.

As will be shown later, the formally exact
fermionic hierarhical equations of motion (HEOM) theory
\cite{Tan89101,Jin08234703},
which has been employed to solve quantum impurity problems
\cite{Li12266403,Zhe13086601,Hou14045141,Har13235426,Sch16201407},
can be established based on \Eq{eom-brhos}.
The analogues of \Eq{eom-brhos} have been obtained
in the forms of stochastic quantum Liouville equation \cite{Hsi18014103}
and non-Markovian QSD equation \cite{Che13052108}.
Making practical use of \Eq{eom-brhos} with conventional
stochastic algorithms faces fundamental difficulties,
which originate from the aforementioned nature of Grassmann variables.
For instance, although Grassmann variables can be represented
by mutually anticommutating matrices,
it would require a huge number of matrices
with huge dimensions to completely model all the time-dependent
AGFs involved in \Eq{eom-brhos}.
Such kind of difficulties has prohibited any direct numerical
application of \Eq{eom-brhos} or its analogues.

For practical purposes, the Grassmann--Wiener processes in \Eq{eom-brhos}
need to be replaced by some operable quantities.
To show such replacement is possible,
let us start with a prototypical equation of motion,
\be
  \dot{y} = y \left[D(t)\,\eta_t
   + \int_{t_0}^t C(\tau)\,\bareta_\tau {\rm d}\tau \right].
\label{eom-y-1}
\ee
While $\eta_t$ and $\bareta_\tau$ are time-dependent AGFs,
$C(t)$ and $D(t)$ are conventional functions.
The stochastic average, $\la y \ra$, is defined similarly to \Eq{rho-tot}.
Like \Eq{eom-brhos}, \Eq{eom-y-1} cannot be solved directly.

We now propose by intuition and will verify analytically later
a mapping scheme, denoted by
\be \label{sub-2}
 \eta_{t} \mapsto v_{t} \Xd, \ \ \bareta_{t} \mapsto v_{t} \Xu.
\ee
By \Eq{sub-2}, each pair of AGFs, $\eta_{t}$ and $\bareta_{t}$,
is mapped to a Gaussian white noise, $v_{t}$,
and a pair of time-independent pseudo-operators,
$\Xu$ and $\Xd$, defined in the space of $S = \{-1,0,1\}$.
Let $\ty = \sum_{l\in S} \ty^{[l]}$. The action of
$X^\pm$ on $\ty^{[l]}$ gives
%
$\ti{y}^{[l]} X^{\pm} = \delta^{\pm}_{l}\, \ti{y}^{[l\pm 1]}$,
%
with $\delta^{\pm}_{0}=\delta^{+}_{-1}=-\delta^{-}_{1}=1$
and $\delta^{-}_{-1}=\delta^{+}_{1}=0$.
This thus transforms \Eq{eom-y-1} into
a normal stochastic differential equation (SDE)
as
\be
  \dot{\tilde{y}} = \ty \left[D(t)\,v_t \Xd
   + \int_{t_0}^t C(\tau)\,v_\tau {\rm d}\tau \Xu \right].
  \label{eom-by-1}
\ee
%
%
%
%
Interestingly, 
one can prove $\la \ty \ra = \mM(\ty^{[0]})$
with $\mM$ denoting the stochastic average exactly
reproduces $\la y \ra$ \cite{SM1}.

In general cases replacing the AGFs
by $\{v_t\}$ and $X^\pm$ is not always exact.
It is noted that \Eqs{def-gt} and \eqref{eom-brhos}
include convolution of memory,
whereas \Eq{eom-y-1} or \Eq{eom-by-1} does not.
The use of the finite space $S$
may cause loss of memory when tracing the cumulative
influence of the AGFs.
Nevertheless, the above example suggests
that it is possible to convert a Grassmann-valued equation to a normal SDE.

We now apply the same strategy to \Eq{eom-brhos} for the AGFs,
\emph{i.e.}, $\eta_{jt} \mapsto v_{jt}\Xd_j$ and
$\bareta_{jt} \mapsto v_{jt}\Xu_j$.
The Gaussian white noises, $\{v_{jt}; j=1,\cdots, 4\}$, satisfy
$\mM(v_{jt}) = 0$ and $\mM(v_{jt}v_{j'\tau})=\delta_{jj'}\delta(t-\tau)$.
The pseudo-operators $X^{\pm}_j$ defined in the space $S_j = \{-1,0,1\}$
are presumed to commute with $\hc$ and $\hc^\dag$.
The resulting formulation can be further simplified by
setting $X^\pm_3 = X^\pm_1$ and $X^\pm_4 = X^\pm_2$
without changing the numerical outcomes.
%
Correspondingly, \Eq{def-gt} is mapped to
\be
\begin{split}
  \tgm_t &= \int_{t_0}^t \left\{
  [C^+(t-\tau)]^\ast v_{4\tau} - iC^-(t-\tau) v_{2\tau} \right\} {\rm d}\tau,
\\
  \tgp_t &=  \int_{t_0}^t \left\{
  [C^-(t-\tau)]^\ast v_{3\tau} - iC^+(t-\tau) v_{1\tau} \right\} {\rm d}\tau.
\label{def-tgt}
\end{split}
\ee
Denote also
\be\label{def-y-1}
\begin{split}
 Y_1 &\equiv v_{1t} \Xd_1 + \tgm_t \Xd_2,
\quad\   Y_2 \equiv v_{2t} \Xu_2 - \tgp_t \Xu_1,
\\
 Y_3 &\equiv v_{3t} \Xd_1 - i \tgm_t \Xd_2,
\ \  \,
 Y_4 \equiv v_{4t} \Xu_2 + i \tgp_t \Xu_1.
\end{split}
\ee
We thus map \Eq{eom-brhos} to a numerically feasible form as
\begin{align}
  \dot{\tilde{\rho}}_{_{\rm S}}
&=  -i [ \Hs, \trhos]
  + e^{-i\pi/4}(\hc^\dag\, Y_1 + Y_2 \,\hc)\trhos
\nl&\quad
  + e^{i\pi/4} \trhos(\hc^\dag \,Y_3 + Y_4\, \hc).
\label{seom-brhos}
\end{align}
Here, $\trhos = \sum_{l_1 \in S_1} \sum_{l_2 \in S_2}
 \trhos^{[l_1 l_2]}$ is defined
in the product space of $V \equiv \Vs \otimes S_1 \otimes S_2$,
with $\Vs$ being the system subspace.
$X^\pm_j$ can act to the left or right of $\trhos$,
and the results of actions
$X^\pm_j \trhos^{[l_1 l_2]} = (-1)^{l_1+l_2} \trhos^{[l_1 l_2]}X^\pm_j$
are given in Supplemental Material \cite{SM1}.
Equation~\eqref{seom-brhos} is to be solved by a
stochastic algorithm, with the initial condition
$\trhos(t_0) = \trhos^{[00]}(t_0) = \rho(t_0)$.
The reduced system density matrix is finally obtained via
\be
 \rho = \la \trhos \ra = \mM \big(\trhos^{[00]}\big).
 \label{trhos-SEOM}
\ee

Equation~\eqref{seom-brhos} is illustrated as
the result of mapping (iii) in \Fig{fig1}.
There, the space $S_j$ is represented by a ladder,
whose three rungs, corresponding to the elements $1$, $0$ and $-1$,
can be interpreted as one-particle, vacuum,
and one-hole pseudo-states, respectively.
The actions of pseudo-operators $X^\pm_j$ result in
transitions between these pseudo-states.
Therefore, 
the mapping (iii) in \Fig{fig1} can be viewed
as a single-configuration-interaction treatment for
the bath, and the system-bath dissipation processes are
modeled by stochastic exchanges of fermion
particles between the system and the pseudo-states.
Moreover, the quantized pseudo-states highlight
the \emph{nonclassical} nature of the AGFs
associated with the quantum fluctuations of fermions.

We now assess whether and how the mapping (iii) in
\Fig{fig1} retains or compromises the exactness of \Eq{eom-brhos}.
This is done by relating \Eq{seom-brhos} to the HEOM formalism.
Traditionally, the HEOM are constructed based on
exponential unravelling of bath correlation functions,
\be\label{ct-exp-1}
  C^\pm(t) = \sum_{m} C^\pm_m(t) = \sum_m  A^\pm_m \,e^{\gamma^\pm_m t},
\ee
subject to the symmetry $\gamma^+_m = (\gamma^-_m)^\ast$
via the fermionic fluctuation-dissipation theorem \cite{Jin08234703}.
In the HEOM theory, a general $(I+J)$th-tier auxiliary
density operator (ADO) is defined in the path-integral form
of ($I$ and $J$ are arbitrary nonnegative integers) \cite{Han18234108}:
\begin{align}
\rho_{m_1 \ldots m_I n_1 \cdots n_J}^{(-\cdots-+\cdots+)} &=
\int \mD \bm\bpsi \mD\bm\psi \mD \bm\bpsi' \mD\bm\psi'\,
 e^{i\mathcal{S}_{f}} \mathcal{F}_{\rm FV}\, e^{-i\mathcal{S}_{b}} \nl
 &\quad \times \mB^{-}_{m_I} \cdots \mB^{-}_{m_1}
\mB^{+}_{n_J} \cdots \mB^{+}_{n_1} \rho(t_0).
\label{ado-heom-1}
\end{align}
Here, $\bm\bpsi = \{\bpsi_\tau\}$ and $\bm\bpsi' = \{\bpsi_\tau'\}$
are Grassmann variables associated with $\hc^\dag$, and
$\bm\psi = \{\psi_\tau\}$ and $\bm\psi' = \{\psi_\tau'\}$
are associated with $\hc$, respectively.
$\mathcal{S}_f$ ($\mathcal{S}_b$) is the forward (backward) action functional
associated with $\Hs$, $\mathcal{F}_{\rm FV}$ is the Feynman--Vernon
influence functional \cite{Fey63118,Jin08234703}, and
\be
\begin{split}
  \mB^{-}_m &=  -i \int_{t_0}^t {\rm d}\tau
   \left[ A^-_m  \, \psi_\tau - \big(A^{+}_m \big)^\ast  \, \psi'_\tau
   \right]  e^{\gamma_m^{-}(t-\tau)},
\\
  \mB^{+}_n &=  -i \int_{t_0}^t {\rm d}\tau
   \left[ A^+_n  \, \bpsi_\tau - \big(A^{-}_n \big)^\ast \, \bpsi'_\tau
   \right]  e^{\gamma_n^{+}(t-\tau)}.  \label{def-b-terms-1}
\end{split}
\ee

It can be proved that, based on unravelling of $g^\pm_t = \sum_m g^\pm_m(t)$
via \Eq{ct-exp-1} and the one-to-one correspondence between $\mB^\pm_m$ and $g^\pm_m$,
the $(I+J)$th-tier ADO is retrieved exactly by
the formal solution of \Eq{eom-brhos} as \cite{SM1}
\be
 \rho^{(-\cdots-+\cdots+)}_{m_1\cdots m_I n_1\cdots n_J}
= e^{i(I+J)\pi/4}
\la g^-_{m_1}\cdots g^-_{m_I} \trhos g^+_{n_1}\cdots g^+_{n_J}\ra.
\label{ado-seom-1}
\ee
In parallel, the $(I+J)$th-tier ADO from \Eq{seom-brhos}
reads
\begin{align}
 \trho^{(-\cdots-+\cdots+)}_{m_1\cdots m_I n_1\cdots n_J}
 &= e^{i(I+J)\pi/4}
  \big\la \tgm_{m_1} \Xd_2 \cdots \tgm_{m_I} \Xd_2 \,\trhos \nl
 &\qquad \qquad \quad \times \tgp_{n_1} \Xu_1 \cdots \tgp_{n_J} \Xu_1 \big\ra.
 \label{ado-seom-2}
\end{align}
%
Such a density variable is automatically zero if it involves two or more identical pseudo-operators $X^\pm_j$.
This is because $\{(X^\pm_j)^p\, \ty\}^{[00]} = \{\ty\,(X^\pm_j)^p \}^{[00]} = 0$
holds for any vector $\ty$ in the space $V$ and $p \geqslant 2$.
In the context of HEOM, this amounts to setting all
\emph{interference} ADOs to zero.
Here, \emph{interference} means the right-hand side of \Eq{ado-heom-1}
involves two or more $\mB^\pm_m$-terms that differ only in $m$.
%
%
%
Consequently, \Eq{seom-brhos} is not equivalent to the exact HEOM,
but correspond to a simplified version of HEOM. 
It thus becomes clear that the substitution of \Eq{sub-2}
is an approximation, which leads to the missing of certain
detailed information on the system dissipative dynamics.

Extension of \Eqs{eom-brhos} and \eqref{seom-brhos} to
general multi-level systems is straightforward,
and the above assessment remains true.
In the case of $\Hsb = \sum_{\nu=1}^{N_\nu} (\hc_\nu^\dag \hF_\nu
+ \hF_\nu^\dag \hc_\nu)$ with $N_\nu$ being the system's
degrees of freedom, the decoupling of system and bath
is realized by introducing the AGFs
$\{\bareta_{j\nu t}\}$ and $\{\eta_{j\nu t}\}$,
which are then represented by Gaussian white noises
$\{v_{j\nu t}\}$ and pseudo-operators $\{X^\pm_{j \nu}\}$
in the same way as \Eq{sub-2}.

The simplified-HEOM (sim-HEOM) method has been established
in Ref.~\onlinecite{Han18234108}, with the role of
interference ADOs discussed extensively therein.
Because of their formal equivalence,
the SEOM of \Eq{seom-brhos} and its multi-level extension
share the same features as the sim-HEOM \cite{Han18234108}:
(a) They yield exact $\rho$ if $C^\pm(t)$ is
a single exponential function. This is obvious because the resulting hierarchy
does not involve any interference ADO.
(b) For general noninteracting systems
they preserve the exact reduced single-electron density matrix $\bm\varrho$
as well as any system property that can be evaluated
from $\bm\varrho$.
This is because the omitted interference ADOs
have no influence on $\bm\varrho$.
(c) For interacting systems they are in principle approximate,
and the interference ADOs are important for the quantitative
description of strong correlation effects such as Kondo phenomena.
Nevertheless, as will be shown below, \Eq{seom-brhos} can still provide
reasonably accurate predictions for system dynamical properties.

\begin{figure}[t]
  \centering
  \includegraphics[width=\columnwidth]{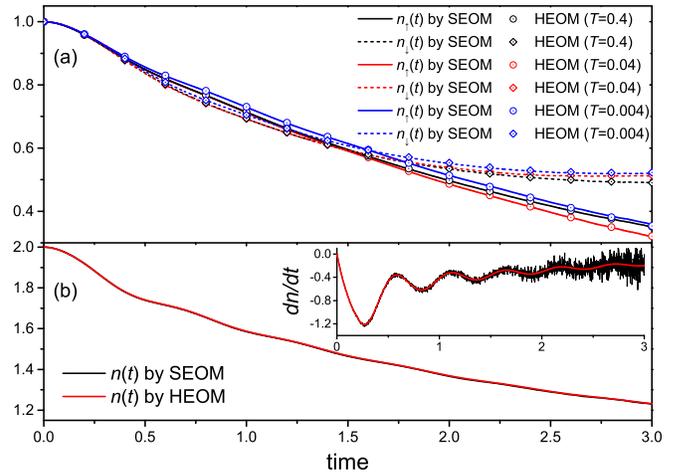}
  \caption{(a) Time evolution of $n_s(t)$ for
  different temperature $T$ calculated with the SEOM and HEOM methods.
  The parameters adopted are (in arbitrary unit):
  $\ep_\uparrow = 0.5$, $\ep_\downarrow = -0.5$,
  $U=5$, $\Gamma=0.5$, $\Omega=0$ and $W = 5$.
  The Euler--Maruyama algorithm \cite{Klo92} is employed to
  solve the SEOM with a time step of $dt = 0.001$.
  The number of trajectories is $N_{\rm traj} = 5\times 10^6$
  for all temperatures.
  (b) Time evolution of $n(t) = n_\uparrow(t) + n_\downarrow(t)$
  calculated with the SEOM and HEOM methods.
  The impurity level energy is shifted by $\Delta \ep = -5.0$
  during the time interval $0.1 < t < 0.2$.
  The other parameters adopted are (in arbitrary unit):
  $\ep_\uparrow = \ep_\downarrow = -2.5$, $U=10$,
  $\Gamma=0.5$, $\Omega=0$, $W = 5$ and $T=0.01$.
  The Euler--Maruyama algorithm is employed with $dt = 0.001$
  and $N_{\rm traj} = 1\times 10^6$. The inset depicts
  ${\rm d}n/{\rm d}t$ versus $t$, which reveals the stochastic error
  of the SEOM in the large-$t$ regime.}
  \label{fig2}
\end{figure}

We now demonstrate the practicality and accuracy of
\Eq{seom-brhos} with open electronic systems
described by the single-impurity Anderson model.
The impurity (system) Hamiltonian is
$\Hs = \sum_{s=\uparrow,\downarrow} \ep_{s}\, \hat{n}_{s}
+ U \hat{n}_\uparrow \hat{n}_\downarrow$,
where $\ep_{s}$ is the energy of spin-$s$ level,
$\hat{n}_{s}$ is the electron number operator,
and $U$ is the electron-electron Coulomb interaction energy.
The reservoir (bath) Hamiltonian  is $\Hb = \sum_{ks} \ep_{ks}\, \hat{n}_{ks}$,
and its influence on the impurity is characterized
by the hybridization functions,
which assume a Lorentzian form of
$\Delta_s(\w) \equiv \pi \sum_k |t_{ks}|^2 \delta(\w -\ep_{ks})
= \frac{\Gamma}{2}\frac{W^2}{(\w - \Omega)^2 + W^2}$.
Here, $\Gamma$, $\Omega$ and $W$ are the effective
impurity-reservoir coupling strength, and the
reservoir band-center and bandwidth, respectively.

Suppose at initial time ($t_0 = 0$) the decoupled impurity
is doubly occupied by spin-up and spin-down electrons.
%
$\Hsb$ is turned on at $t > 0$,
which triggers the electron transfer
between the impurity and reservoir.
The time evolution of $\rho(t)$ is obtained by solving
a spin-resolved version of \Eq{seom-brhos}.
The number of electrons on the impurity
is computed by $n_s(t) = \trs[\hat{n}_{s} \rho(t)]$
and shown in \Fig{fig2} along with
quantitatively accurate results obtained by the full HEOM
(with all ADOs kept).
For all the systems examined in \Fig{fig2},
$U$ assumes an appreciable value,
yet the results of our proposed SEOM agree
remarkably with the HEOM counterparts.

Regarding numerical efficiency,
the SEOM does not require an explicit unravelling of
$C^\pm(t)$, and hence its memory cost is
substantially smaller than the HEOM.
This allows the SEOM to exploit the regime
of extremely low temperatures which remains
prohibitive for the present HEOM.
Moreover, the trajectory-based algorithms for
the SEOM could benefit from the
massive parallel computing techniques.

Admittedly, there may exist some strongly correlated
quantum impurity systems, for which the
substitution of \Eq{sub-2} for the AGFs
and the resulting SEOM lead to less satisfactory
numerical descriptions.
Even for such systems, the proposed SEOM
still lays a valuable foundation
for future development of more sophisticated practical schemes.
For instance, the two-electron interaction in $\Hs$
can be equivalently characterized by interactions between
each electron and auxiliary stochastic fields via a Hubbard--Stratonovich
transformation \cite{Hub5977},
and thus the system becomes effectively noninteracting.
The resulting SEOM is expected to yield the exact $\bm\varrho$
for the effective noninteracting system,
from which any physical observable of the
original interacting dissipative system can be evaluated.
%


Support from the Ministry of Science and Technology of China
(Grants No.\ 2016YFA0400900 and No.\ 2016YFA0200600),
the National Natural Science Foundation of China
(Grants No.\ 21573202, No.\ 21633006 and No.\ 21373064),
the MOE of China (111 Project Grant No.\ T21052018003),
the Fundamental Research Funds for the Central Universities
(Grant No.\ 2340000074), and the SuperComputing Center of USTC is
gratefully acknowledged.


\end{document}


\title{Stochastic Equation of Motion for Dissipative Dynamics of Fermionic Open Systems}

\author{Lu Han}
\affiliation{Hefei National Laboratory for Physical Sciences at the Microscale \&
Synergetic Innovation Center of Quantum Information and Quantum Physics \&
CAS Center for Excellence in Nanoscience,
University of Science and Technology of China, Hefei, Anhui 230026, China}

\author{Vladimir Chernyak}
\affiliation{Hefei National Laboratory for Physical Sciences at the Microscale \&
Synergetic Innovation Center of Quantum Information and Quantum Physics \&
CAS Center for Excellence in Nanoscience,
University of Science and Technology of China, Hefei, Anhui 230026, China}
\affiliation{Department of Chemistry, Wayne State University, 5101 Cass Avenue, Detroit, MI 48202}

\author{Yun-An Yan}
\affiliation{Guizhou Provincial Key Laboratory of Computational Nano-Material Science,
Guizhou Normal College, Guizhou 550018, China}

\author{Xiao Zheng} \email{xz58@ustc.edu.cn}
\affiliation{Hefei National Laboratory for Physical Sciences at the Microscale \&
Synergetic Innovation Center of Quantum Information and Quantum Physics \&
CAS Center for Excellence in Nanoscience,
University of Science and Technology of China, Hefei, Anhui 230026, China}

\author{YiJing Yan}
\affiliation{Hefei National Laboratory for Physical Sciences at the
Microscale \& \iChEM, University of Science and Technology of China, Hefei, Anhui
230026, China}

\date{\today}

\noindent {\bf Supplemental Materials for}

\maketitle

\tableofcontents
\clearpage

\renewcommand{\theequation}{S\arabic{equation}}
\renewcommand{\thetable}{S\arabic{table}}
\renewcommand{\thefigure}{S\arabic{figure}}

%

\section{Decoupling the dynamics of system and bath}

Based on the Gaussian integral for Grassmann variables, we have the equality
\be
  e^{A \bar{\psi}\theta} = \int d\bareta d\eta \, e^{-\bareta\eta + \sqrt{A} \bar{\psi}\eta + \sqrt{A} \bareta\theta}.  \label{hs-0}
\ee
Here, $A$ is a c-number, and $\{\bareta, \eta, \bpsi, \theta\}$ are Grassmann variables
which anticommute with each other.
%

%
Consider a single-level system coupled to a fermion bath. 
%
In the fermionic coherent-state path-integral representation, the forward propagator of
the system-bath interaction Hamiltonian is (we set $\hbar = 1$ hereafter)
\begin{align}
  \mathcal{U}_{\rm SB}(t,t_0) &= \exp_{+}\,\left\{-i\int_{t_0}^t d\tau\,  \Hsb \right\} \nl
  &= \int_{t_0}^t \mD \bm\bpsi \, \mD \bm\psi\, \mD \bm\btheta\, \mD \bm\theta \,
  e^{-i \int_{t_0}^t  (\bpsi_\tau \theta_\tau + \btheta_\tau \psi_\tau) d\tau } \nl
  &= \int_{t_0}^t \mD \bm\bareta_1  \mD \bm\eta_1 \mD \bm\bareta_2 \mD \bm\eta_2\,
  e^{- \int_{t_0}^t (\bareta_{1\tau} \eta_{1\tau} - \bareta_{2\tau} \eta_{2\tau})\, d\tau} \nl
  &\qquad \times
  \mD \bm\bpsi\, \mD \bm\psi \, \mD \bm\btheta \, \mD\bm\theta
  \, e^{\int_{t_0}^t  B  (\bpsi_\tau \eta_{1\tau} + \bareta_{2\tau}\psi_\tau) d\tau }
  \, e^{\int_{t_0}^t  B  (\btheta_\tau \eta_{2\tau} + \bareta_{1\tau}\theta_\tau) d\tau }.
  \label{usb-1}
\end{align}
Here, $B = e^{-i\pi/4}$, and
$\{\bm\bareta_{j}, \bm\eta_j\} = \{ \bareta_{j\tau}, \eta_{j\tau}\}$
for $t_0 < \tau < t$.
The time-dependent Grassmann variables $\{\bpsi_\tau,\psi_\tau,\btheta_\tau,\theta_\tau\}$
are associated with the operators $\{\lamh \hc^\dag, \lamh \hc , \lammh \hF^\dag, \lammh \hF \}$
in the path-integral formulation, with $\lambda$ being a reference energy of any positive value.
For simplicity, we choose $\lambda = 1$ in the main text and throughout this Supplemental
Material.

The backward propagator $\mathcal{U}_{\rm SB}^\dag(t,t_0)$ can be expressed similarly by introducing
the auxiliary Grassmann fields (AGFs) $\{\bareta_{3\tau}, \eta_{3\tau}, \bareta_{4\tau}, \eta_{4\tau}\}$.
%
%
%
With the use of AGFs, the system and bath is formally decoupled from each other.
Instead, they are coupled to the AGFs $\{\bareta_{j\tau}, \eta_{j\tau}\}$
($j=1,\ldots,4$). With the initial factorization condition of
$\rhot(t_0) = \rhos(t_0)  \rhob(t_0)$, the equations of motion (EOM)
for $\rhos$ and $\rhob$ are given by Eqs.~(1) and (2) in the main text.
%
%
The density matrix of the total system is obtained by 
$\rhot = \la \rhos  \rhob \ra$.
%
%
From the It\^{o}'s formula, we have
\begin{align}
  d\rhot &= \left \la  (d\rhos) \rhob + \rhos (d\rhob) + (d\rhos) (d\rhob)  \right\ra \nl
  &= -i [\Hs + \Hb, \rhot] \,dt
  - i \,\big\la (\hc^\dag \eta_1 + \bar{\eta}_2 \hc ) \rhos  (\bar{\eta}_1 \hF + \hF^\dag \eta_2 ) \rhob \big\ra
  \, (dt)^2  \nl
  &\quad + i\, \big\la \rhos (\hc^\dag \eta_3 + \bar{\eta}_4 \hc )  \rhob
   (\bar{\eta}_3 \hF + \hF^\dag \eta_4 ) \big\ra\, (dt)^2  \nl
  &= -i [\Hs + \Hb, \rhot] \,dt  -i \,\big\la (\hc^\dag \eta_1 + \bar{\eta}_2 \hc) (\bar{\eta}_1 \hF + \hF^\dag \eta_2)
  \big\ra \la \rhos  \rhob \ra(dt)^2  \nl
  &\quad  + i\, \la \rhos  \rhob \ra \big\la (\hc^\dag \eta_3 + \bar{\eta}_4 \hc ) (\bar{\eta}_3 \hF + \hF^\dag \eta_4 ) \big\ra\,
   (dt)^2  \nl
  &= -i [\Hs + \Hb + \Hsb, \rhot]\, dt \nl
  &= -i [H_{_{\rm T}}, \rhot]\, dt.  \label{check-eom-tot}
\end{align}
Here, we have used the causality relation that $\rhos$ and $\rhob$ at time $t$
depend only on AGFs at time $\tau < t$;
the equalities $\la \bareta_{j\tau} \ra = \la \eta_{j\tau} \ra = 0$ and
$\la \eta_{j\tau} \bareta_{j'\tau'} \ra = \delta_{jj'}\delta(\tau-\tau')$;
together with the equalities
$\rhos (\bar{\eta}_1 \hF + \hF^\dag \eta_2) = (\bar{\eta}_1 \hF + \hF^\dag \eta_2)\rhos$
and $(\hc^\dag \eta_3 + \bar{\eta}_4 \hc)\rhob =\rhob (\hc^\dag \eta_3 + \bar{\eta}_4 \hc)$.
%


In the $\Hb$--interaction picture, define
the bath density matrix
$\trhob \equiv e^{i\int_{t_0}^t \! \Hb d\tau} \rhob e^{-i\int_{t_0}^t \! \Hb d\tau}$
and the operator
$\hF(t) \equiv e^{i\int_{t_0}^t \!\Hb d\tau} \hF \, e^{-i\int_{t_0}^t \!\Hb d\tau}$.
From Eq.~(2) of main text, we have
\be
  \dot{\tilde{\rho}}_{_{\rm B}} = B\, \big[\bareta_{1t} \hF(t) + \hF^\dag(t)\, \eta_{2t}\big]\,\trhob
 + B^\ast \trhob \big[\bareta_{3t}\hF(t) + \hF^\dag(t)\, \eta_{4t}\big], \label{eom-brhob}
\ee
which can be solved formally by the Magnus expansion.
If the bath is initially in thermal equilibrium, and $\rhob(t_0) = \rhob^{\rm eq}$
satisfies Gaussian statistics, we have
\be
 \trb(\rhob) = \trb(\trhob) = e^{\int_{t_0}^t
 \left[(\bareta_{1\tau} - i\bareta_{3\tau}) \gm_\tau +
 (\eta_{2\tau} - i\eta_{4\tau}) \gp_\tau \right] d\tau}.
 \label{tr-rhob}
\ee
\section{Formal equivalence between Eq.~(5) of main text 
and the fermionic HEOM formulation}

The fermionic hierarchical equations of motion (HEOM) are constructed based
on unravelling of two-time bath correlation functions by exponential functions:
%
$C^\sigma(t) = \sum_m C_m^\sigma(t) = \sum_m A_m^\sigma \,e^{\gamma_m^\sigma t}$
with $\sigma=+$ or $-$.
In the HEOM theory, the EOM for $\rho_{m_1 \ldots m_I n_1 \cdots n_J}^{(-\cdots-+\cdots+)}$
can be recast into a compact form of
\begin{align}
 \dot{\rho}_{m_1 \ldots m_I n_1 \cdots n_J}^{(-\cdots-+\cdots+)} &=
 \Big(-i \mathcal{L}_{_{\rm S}} + \sum_{i=1}^I \gamma_{m_i}^-
 + \sum_{j=1}^J \gamma_{n_j}^+ \Big)\, \rho_{m_1 \ldots m_I n_1 \cdots n_J}^{(-\cdots-+\cdots+)}
 + \sum_{i=1}^I  \mathcal{C}^-_{m_i}\, \rho_{m_1 \ldots m_{i-1} m_{i+1} \cdots n_J}^{(-\cdots--\cdots+)} \nl
 &\quad + \sum_{j=1}^J  \mathcal{C}^+_{n_j} \, \rho_{m_1 \ldots n_{j-1} n_{j+1} \cdots n_J}^{(-\cdots++\cdots+)}
 + \sum_{\sigma=+,-} \sum_{r}  \mathcal{A}^\sigma_r \, \rho_{m_1 \ldots m_I r n_1 \cdots n_J}^{(-\cdots-\sigma+\cdots+)},
 \label{heom-1}
\end{align}
where $\mathcal{L}_{_{\rm S}}\star \equiv [\Hs, \star]$.
The detailed forms of the superoperators
$\{\mathcal{C}^-_{m_i}, \mathcal{C}^+_{n_j}, \mathcal{A}^\sigma_r\}$
have been given by Eqs.~(26)--(29) in Ref.~\onlinecite{Han18JCP}.

Regarding the Grassmann-valued SEOM for $\trhos$ given by Eq.~(5) in the main text, the bath-induced
AGFs are decomposed as $g^\sigma_t = \sum_m g^\sigma_m(t)$.
%
%
The EOM for each component is self-closed:
\begin{align}
   \dot{g}^-_m &= \left[ -i A_m^{-}\, \eta_{2t} + (A_m^{+})^\ast \,\eta_{4t} \right]
   + \gamma_m^{-}\, g^-_m, \nl
   \dot{g}^+_n &= \left[ -i A_n^{+}\, \bar{\eta}_{1t} + (A_n^{-})^\ast\, \bar{\eta}_{3t} \right]
   + \gamma_m^{+}\, g^+_n.
\end{align}
The same $(I+J)$th-tier ADO can be retrieved by $\trhos$ of Eq.~(5) in the main text as follows:
\be
 \rho_{m_1 \ldots m_I n_1 \cdots n_J}^{(-\cdots-+\cdots+)} \equiv
(B^\ast)^{I+J}\, \la \gm_{m_1} \cdots \gm_{m_I} \trhos \,\gp_{n_1} \cdots \gp_{n_J} \ra.
\label{ado-seom-1}
\ee
Based on It\^{o}'s formula, its differential consists of three parts:
\be
  d \rho_{m_1 \cdots m_I n_1 \cdots n_J}^{(-\cdots-+\cdots+)} = \Xi_1 + \Xi_2 + \Xi_3.
  \label{heom-ado-1}
\ee
Presuming the system creation and annihilation operators ($\hc$ and $\hc^\dag$)
\emph{commute} with all the AGFs, and using the equality
\be
\la f(t) \, g_m^\sigma(t) \ra = - \la g_m^{\sigma}(t) f(t) \ra, \label{fg-1}
\ee
which holds for any analytic function $f(t)$ of the AGFs $\{\bar{\eta}_{i\tau},\eta_{i\tau}\}$,
we express the three parts of
$d\rho_{m_1 \cdots m_I n_1 \cdots n_J}^{(-\cdots-+\cdots+)}$ respectively as follows.
\begin{align}
 \Xi_1 &= (B^\ast)^{I+J} \la \gm_{m_1} \cdots \gm_{m_I} d\trhos \,\gp_{n_1} \cdots \gp_{n_J} \ra \nl
  &= -i \big[ \Hs, \rho_{m_1 \ldots m_I n_1 \cdots n_J}^{(-\cdots-+\cdots+)} \big] dt
 -i \,(B^\ast)^{I+J+1} \sum_r  \bigg( \hc^\dag \, \la \cdots \gm_{m_I} \gm_r \trhos \cdots \ra
  +  \la \cdots \trhos \gm_r \gp_{n_1} \cdots \ra \, \hc^\dag  \nl
  &\qquad - \hc\, \la \cdots \gm_{m_I} \gp_r \trhos \cdots \ra
  - \la \cdots \trhos\, \gp_r \gp_{n_1} \cdots \ra \, \hc \bigg) dt \nl
  &=  -i \big[ \Hs, \rho_{m_1 \ldots m_I n_1 \cdots n_J}^{(-\cdots-+\cdots+)} \big] dt
  -i \sum_r \, \bigg( \hc^\dag \, \rho_{m_1 \ldots m_I r n_1 \cdots n_J}^{(-\cdots--+\cdots+)}
  -(-1)^{I+J} \rho_{m_1 \ldots m_I r n_1 \cdots n_J}^{(-\cdots--+\cdots+)} \hc^\dag \nl
 &\qquad + (-1)^{I+J}  \hc\, \rho_{m_1 \ldots m_I r n_1 \cdots n_J}^{(-\cdots-++\cdots+)}
  - \rho_{m_1 \ldots m_I r n_1 \cdots n_J}^{(-\cdots-++\cdots+)} \,\hc  \bigg) dt.
  \label{term-1}
\end{align}
The causality relation ensures $\trhos(t)$ and $\{g^\sigma_m(t)\}$
depend only on the AGFs prior to the time $t$, and thus we have
\begin{align}
  \Xi_2 &=  \sum_{i=1}^I \la \cdots d\gm_{m_i} \cdots \trhos \cdots \ra +
  \sum_{j=1}^J \la \cdots \trhos \cdots  d\gp_{n_j} \cdots \ra
  =  \Big( \sum_{i=1}^I \gamma_{m_i}^{-} + \sum_{j=1}^J \gamma_{n_j}^{+} \Big)
 \rho_{m_1 \ldots m_I n_1 \cdots n_J}^{(-\cdots-+\cdots+)} \, dt.
 \label{term-2} \\
%
 \Xi_3  &= \sum_{i=1}^I \la \cdots d\gm_{m_i} \cdots d\trhos \cdots \ra +
  \sum_{j=1}^J \la \cdots d\trhos \cdots  d\gp_{n_j} \cdots \ra \nl
 &= (B^\ast)^{I+J-1}  \bigg\{ \sum_{i=1}^I -i A_{m_i}^{-}
  \hc \, \la \cdots \gm_{m_{i-1}} \eta_{2t} \gm_{m_{i+1}} \cdots \bar{\eta}_{2t} \trhos \cdots \ra 
 + i (A^{+}_{m_i})^\ast \la \cdots \gm_{m_{i-1}} \eta_{4t} \gm_{m_{i+1}} \cdots
 \trhos \bar{\eta}_{4t} \cdots \ra \, \hc \nl
 &\quad + \sum_{j=1}^J  -i A_{n_j}^{+} \hc^\dag
 \, \la \cdots \eta_{1t} \trhos \cdots \gp_{n_{j-1}} \bar{\eta}_{1t} \gp_{n_{j+1}} \cdots \ra 
 + i (A^{-}_{n_j})^\ast \la \cdots \trhos \eta_{3t} \cdots \gp_{n_{j-1}} \bar{\eta}_{3t}
 \gp_{n_{j+1}} \cdots \ra \, \hc^\dag \bigg\} \, (dt)^2 \nl
 &= -i \sum_{i=1}^I \left( A_{m_i}^{-} (-1)^{I-i} \, \hc \,
 \rho_{m_1\ldots m_{i-1} m_{i+1}\cdots n_J}^{(-\cdots--\cdots+)}
 - (A^{+}_{m_i})^\ast (-1)^{i-1+J}\, \rho_{m_1\ldots m_{i-1} m_{i+1}\cdots n_J}^{(-\cdots--\cdots+)} \, \hc \right) dt \nl
 &\quad -i \sum_{j=1}^J \left( A_{n_j}^{+}\, (-1)^{I+J-j} \, \hc^\dag \,
 \rho_{m_1\ldots n_{j-1} n_{j+1}\cdots n_J}^{(-\cdots++\cdots+)}
 - (A^{-}_{n_j})^\ast \, (-1)^{j-1}\,
 \rho_{m_1\ldots n_{j-1} n_{j+1}\cdots n_J}^{(-\cdots++\cdots+)} \, \hc^\dag \right) dt.
 \label{term-3}
\end{align}
Here, we have used the equalities $B B^\ast = 1$ and $(B^\ast)^2 = i$, and
\begin{align}
\la \cdots \gm_{m_{i-1}} \eta_{2t} \gm_{m_{i+1}} \cdots \bar{\eta}_{2t} \trhos \cdots \ra dt
 &= (-1)^{I-i} \, \la \eta_{2t} \bareta_{2t} \ra \, \la \cdots \gm_{m_{i-1}} \gm_{m_{i+1}} \cdots \trhos \cdots \ra dt \nl
 &= (-1)^{I-i}  \, \rho_{m_1\ldots m_{i-1} m_{i+1}\cdots n_J}^{(-\cdots--\cdots+)}.
\end{align}
Apparently, the EOM for $\rho_{m_1 \cdots m_I n_1 \cdots n_J}^{(-\cdots-+\cdots+)}$
defined by \Eq{ado-seom-1} is formally identical to \Eq{heom-1}.

\section{Equivalence between the solutions of Eq.~(6) and Eq.~(8) of main text}

We discretize the time domain by setting $t_0 = 0$ and $t = N_t\,dt$, with
$dt$ being the infinitesimal increment time step and $N_t$ the number of steps.
At $t_i = i dt$ and $t_{i+1} = (i+1) dt$, Eq.~(6) of the main text leads to
\begin{align}
   dy_i &= y_i D_i \eta_i + y_i \left(C_0 \bareta_0 + C_1 \bareta_1 + \cdots  + C_{i-1} \bareta_{i-1}+ C_{i} \bareta_{i} \right),
   \label{dyi} \\
   dy_{i+1} &= y_{i+1} D_{i+1} \eta_{i+1} + y_{i+1}
   \left(C_0 \bareta_0 + C_1 \bareta_1 + \cdots + C_{i} \bareta_{i} + C_{i+1} \bareta_{i+1}\right).
   \label{dyi+1}
\end{align}
By causality, $dy_i$ and $y_{i+1}$ depend on $\{\eta_0, \cdots, \eta_i; \bareta_0, \cdots, \bareta_{i}\}$.
Taking the average over Grassmann fields $\{\bareta_j, \eta_j\}$ ($0 \leqslant j \leqslant i+1$)
for both sides of \Eq{dyi+1}, we have
\begin{align}
  \la dy_{i+1} \ra &= \la y_{i+1} D_{i+1} \eta_{i+1} \ra + \la (y_i + dy_i)
  \left(C_0 \bareta_0 + \cdots + C_{i-1} \bareta_{i-1} + C_{i} \bareta_{i} + C_{i+1} \bareta_{i+1}\right) \ra \nl
  &= \la y_i \left(C_0 \bareta_0 + \cdots + C_{i-1} \bareta_{i-1}  + C_{i} \bareta_{i}\right) \ra
  + \la y_i C_{i+1} \bareta_{i+1} \ra +
  \la dy_i \left(C_0 \bareta_0 + \cdots + C_{i-1} \bareta_{i-1} + C_{i} \bareta_{i} + C_{i+1} \bareta_{i+1}\right) \ra \nl
  &= \la dy_i - y_i D_i \eta_i \ra + \la y_i C_{i+1} \bareta_{i+1} \ra +
  \la y_i D_i \eta_i \left(C_0 \bareta_0 + \cdots + C_{i-1} \bareta_{i-1} + C_{i} \bareta_{i}  + C_{i+1} \bareta_{i+1}\right) \ra \nl
  &\quad + \la y_i \left(C_0 \bareta_0 + C_1 \bareta_1 + \cdots + C_{i} \bareta_{i} \right)
  \left(C_0 \bareta_0 + C_1 \bareta_1  + \cdots + C_{i} \bareta_{i}+ C_{i+1} \bareta_{i+1}\right) \ra \nl
  &= \la dy_i \ra +  D_i C_i \la y_i \ra.  \label{recur-1}
\end{align}
Here, the last equality makes use of the causality relation that $y_i$ is independent of
$\eta_i$ and $\bareta_i$.
Equation~\eqref{recur-1} thus gives the formal solution of $\la y \ra$
in the form of recursive relation for $\la dy_i \ra$ at discretized time steps.

Now, with the mapping $\eta_i \mapsto v_i \Xd$ and $\bareta_i \mapsto v_i \Xu$,
$\ty = \sum_{l\in\{-1,0,1\}} \ty^{[l]}$.
\Eqs{dyi} and \eqref{dyi+1} are replaced by
\begin{align}
   d\ty_i &= \ty_i D_i v_i \Xd + \ty_i (C_0 v_0 + \cdots + C_{i-1} v_{i-1}+ C_{i} v_{i}) \Xu,
   \label{dtyi} \\
   d\ty_{i+1} &= \ty_{i+1} D_{i+1} v_{i+1} \Xd +
   \ty_{i+1} (C_0 v_0 + \cdots + C_{i-1} v_{i-1} + C_{i} v_{i} + C_{i+1} v_{i+1}) \Xu.
   \label{dtyi+1}
\end{align}
We have
\begin{align}
 \la d\ty_{i+1} \ra &= \la \ty_{i+1} D_{i+1} v_{i+1} \Xd \ra +
   \la (\ty_{i} + d\ty_i) (C_0 v_0 + \cdots + C_{i-1} v_{i-1} + C_{i} v_{i}+ C_{i+1} v_{i+1}) \Xu \ra \nl
   &= \la d\ty_i - \ty_i D_i v_i \Xd + \ty_i C_{i+1} v_{i+1} \Xu \ra
   + \la d\ty_i (C_0 v_0 + \cdots + C_{i-1} v_{i-1} + C_{i} v_{i}+ C_{i+1} v_{i+1}) \Xu \ra \nl
   &= \la d\ty_i \ra + \la \ty_i D_i v_i \Xd (C_0 v_0 + \cdots + C_{i-1} v_{i-1} + C_{i} v_{i}+ C_{i+1} v_{i+1}) \Xu \ra \nl
   &\quad + \la  \ty_i (C_0 v_0 + \cdots + C_{i-1} v_{i-1}+ C_{i} v_{i}) \Xu (C_0 v_0 + \cdots + C_{i-1} v_{i-1} + C_{i} v_{i} + C_{i+1} v_{i+1}) \Xu \ra \nl
   &= \la d\ty_i \ra + D_i C_i \la \ty_i \ra.  \label{recur-2}
\end{align}
Here, $\la \ty \ra = \mM(\ty^{[0]})$.
To achieve the last equality of \Eq{recur-2}, we need to have
$\la \ty \Xd \Xu \ra = \la \ty \ra$ and $\la \ty (\Xu)^2 \ra = 0$.
These can be easily satisfied, \emph{e.g.}, by setting
\begin{align}
   & \ty_i^{[-1]} \Xd = 0,
                      &\, & \ty_i^{[-1]}\Xu = \ty_i^{[0]}, \nl
   & \ty_i^{[0]} \Xd  = \ty_i^{[-1]},
                      &\, & \ty_i^{[0]} \Xu = \ty_i^{[1]}, \nl
   & \ty_i^{[1]} \Xd  = -\ty_i^{[0]},
                      &\, & \ty_i^{[1]} \Xu = 0.  \label{ladder-ty-1}
\end{align}
Apparently, \Eq{recur-2} is identical to \Eq{recur-1}. Therefore,
the solution of Eq.~(8) in the main text exactly retrieves that of Eq.~(6).
%

%

\section{Remarks on the ladder pseudo-operators}

In the context of Eq.~(10) of the main text,
the ladder pseudo-operators $X^\pm_1$ and $X^\pm_2$
can act to the left and right of $\trhos$, which yield
\begin{align}
  \Xu_1 \,\trhos^{[l_1,l_2]} &= (-1)^{l_1+l_2} \,\trhos^{[l_1,l_2]} \Xu_1
      = \chi^{l_1}_{\{-1,0\}} (-1)^{l_1} \,\trhos^{[l_1+1,l_2]}, \nl
  \Xd_1 \,\trhos^{[l_1,l_2]} &= (-1)^{l_1+l_2} \,\trhos^{[l_1,l_2]} \Xd_1
      = \chi^{l_1}_{\{0,1\}} \,\trhos^{[l_1-1,l_2]}, \nl
  \Xu_2 \,\trhos^{[l_1,l_2]} &= (-1)^{l_1+l_2} \,\trhos^{[l_1,l_2]} \Xu_2
      = \chi^{l_2}_{\{-1,0\}} (-1)^{l_1+l_2} \,\trhos^{[l_1,l_2+1]}, \nl
  \Xd_2 \,\trhos^{[l_1,l_2]} &= (-1)^{l_1+l_2} \,\trhos^{[l_1,l_2]} \Xd_2
      = \chi^{l_2}_{\{0,1\}} (-1)^{l_1} \,\trhos^{[l_1,l_2-1]}.
      \label{ladder-2}
\end{align}
Here, $\chi^{l_j}_{\mathbb{A}} = 1$ (if $l_j\in \mathbb{A}$) or $0$
(if $l_j\notin \mathbb{A}$) is a step function,
which ensures the action of $X^\sigma_j$ does not exceed the
boundary of ladder space $S_j$.
In the following, we elaborate more on the
construction of these pseudo-operators.

Each pseudo-operator can be associated with a time-independent
Grassmann variable, \emph{i.e.},
\be
  \Xu_j  \mapsto \bxi_j, \quad  \Xd_j \mapsto \xi_j.  
\ee
%
A one-to-one mapping can be established between the ladder pseudo-states
$[l_1, l_2]$ and the normal-ordered monomials of Grassmann variables
$\{\bxi_1, \xi_1, \bxi_2, \xi_2\}$.
Here, a monomial is considered to be in \emph{normal} order if
its constituent Grassmann variables follow the sequence of
$\xi_1 \bxi_1 \xi_2 \bxi_2$.
For instance, $\xi_1 \xi_2$ and $\bxi_1 \xi_2$ are in normal order,
while $\bxi_2 \bxi_1$ and $\xi_2\xi_1$ are not.
This means that any vector $f$ in the space $V=V_{_{\rm S}}\otimes S_1 \otimes S_2$
can be represented uniquely by a polynomial of
$\{\bxi_1, \xi_1, \bxi_2, \xi_2\}$ as follows,
\be
  f = \sum_{l_1 \in\, S_1}\sum_{l_2 \in\, S_2}\, f^{[l_1,l_2]}
    \mapsto \sum_{p_1, p_2, p_3, p_4\in \{0,1\}} B_{p_1 p_2 p_3 p_4}
    \, \xi_1^{p_1}\, \bxi_1^{p_2}\, \xi_2^{p_3}\, \bxi_2^{p_4}.
    \label{map-1}
\ee
The ladder space $S_j$ is spanned by only three pseudo-states.
Specifically, $l_j = -1$, $0$, and $1$ correspond to
$\xi_j$, $1$, and $\bxi_j$, respectively;
whereas there is no pseudo-state representing the
dual variables $\xi_j \bxi_j$.
Consequently, all monomials involving dual variables
are suppressed in the polynomial of \Eq{map-1},
\emph{i.e.}, $B_{p_1 p_2 p_3 p_4} = 0$ if $p_1 = p_2 = 1$ or
$p_3 = p_4 = 1$.
%

Take the first line of \Eq{ladder-2} as an example ---
the action of $\Xu_1$ to the left of $f^{[l_1,l_2]}$ gives
\be
  \Xu_1 f^{[l_1,l_2]} = \chi^{l_1}_{\{-1,0\}} (-1)^{l_1} f^{[l_1+1,l_2]}.
  \label{Xu1-f-1}
\ee
Here, the step function $\chi^{l_1}_{\{-1,0\}}$ enforces the action
returns zero if $(l_1 + 1)$ exceeds the upper bound of the ladder.
Suppose the pseudo-states $[l_1,l_2]$ and $[l_1+1,l_2]$ correspond
to the normal-ordered monomials 1 and 2, respectively.
The left action of $\Xu_1$ amounts to multiplying $\bxi_1$
to the left of the monomial 1, which results in the monomial $1'$
(with dual variables suppressed).
The prefactor $(-1)^{l_1} = 1$ (or $-1$) indicates that
it requires an even (or odd) number of swaps of Grassmann variables
to rearrange the monomial $1'$ into the normal-ordered monomial $2$.

For instance, the pseudo-state $[l_1,l_2] = [-1,1]$
corresponds to the normal-ordered monomial $\xi_1 \bxi_2$,
and $[l_1+1,l_2] = [0,1]$ corresponds to the monomial $\bxi_2$.
Multiplying $\bxi_1$ to the left of $\xi_1 \bxi_2$ yields
$\bxi_1 \xi_1 \bxi_2 = -\xi_1 \bxi_1 \bxi_2 \mapsto -\bxi_2$.
In the last step, the dual term $\xi_1 \bxi_1$ is suppressed
(reduced to $1$),
and the resulting minus sign recovers the prefactor $(-1)^{l_1} = -1$
in \Eq{Xu1-f-1}.

\section{Formal equivalence between Eq.~(11) of main text
and the simplified-HEOM formulation}

In relation to \Eq{ado-seom-1}, with the AGFs represented
by Gaussian white noises and ladder pseudo-operators, the $(I+J)$th-tier
ADO is constructed by
\be
\trho_{m_1 \ldots m_I n_1 \cdots n_J}^{(-\cdots-+\cdots+)}
= (B^\ast)^{I+J} \la \tgm_{m_1} \Xd_2 \cdots \tgm_{m_I} \Xd_2 \brhos
\tgp_{n_1} \Xu_1 \cdots \tgp_{n_J} \Xu_1 \ra,
\label{ado-seom-2}
\ee
where $\{\tgm_m\}$ and $\{\tgp_n\}$ are bath-induced stochastic fields
\begin{align}
  \tgm_m(t) &= \int_{t_0}^t
  \left[ -i A^{-}_m \,v_{2\tau} + \big(A^{+}_m \big)^\ast \,v_{4\tau} \right] e^{\gamma_m^{-}(t-\tau)}\, d\tau,  \nl
  \tgp_n(t) &= \int_{t_0}^t
  \left[ -i A^{+}_n\, v_{1\tau} + \big(A^{-}_n \big)^\ast\, v_{3\tau} \right] e^{\gamma_n^{+}(t-\tau)}\, d\tau.
\end{align}
For each ladder space $S_j$, only one pseudo-operator
($\Xu_j$ or $\Xd_j$) is involved in the construction of ADOs,
and pseudo-operators belonging to different ladder spaces
anticommute with each other.

Because of the finite dimension of $S_j$,
$\la (X^\sigma_j)^p f \ra = \la f (X^\sigma_j)^p \ra = 0$ holds
for any $f = \sum_{l_1\in S_1}\sum_{l_2\in S_2} f^{[l_1,l_2]}$ and $p \geqslant 2$.
Regarding \Eq{ado-seom-2}, it is immediately recognized
that the ADO is zero if the right-hand side involves two or more
identical $X^\sigma_j$.
In the context of original HEOM, this amounts to setting
any ADO that involves two or more $\mB^\sigma_m$--terms
that differ only in the index $m$ to zero;
see Eq.~(15) of main text.
Such ADOs are referred to as \emph{interference} ADOs,
which are important for the accurate description of
strongly correlated states in fermionic dissipative systems.

By using the property that $X^\pm_j$ commute with $\hc$ and $\hc^\dag$,
as well as the equality
\be
\la X^\sigma_j f \ra = -\la f X^\sigma_j \ra, \label{fX-1}
\ee
which is in parallel with \Eq{fg-1},
the time differential of any nonzero
$\trho_{m_1 \ldots m_I n_1 \cdots n_J}^{(-\cdots-+\cdots+)}$
is also given by \Eqs{term-1}, \eqref{term-2} and \eqref{term-3}.
Therefore, Eq.~(11) of main text is formally equivalent to
the simplified-HEOM (sim-HEOM) formulation
in which all \emph{interference} ADOs are omitted from
the original HEOM.
The detailed derivation and important properties of
the sim-HEOM method have been elaborated in Ref.~\onlinecite{Han18JCP}.

In general cases where a multi-level system
is coupled to more than one fermion baths, the
%
$\tilde{g}^\sigma_m X^\sigma_j$ in \Eq{ado-seom-2} is replaced
by $\tilde{g}^\sigma_{\nu\alpha m} X^\sigma_{j\nu\alpha}$,
where $\nu$ labels the system levels and $\alpha$ labels the baths.
Correspondingly, $\mB^\sigma_m$ is replaced by $\mB^\sigma_{\nu\alpha m}$.
Again, \emph{interference} means the ADO involves two or more
$\mB^\sigma_{\nu\alpha m}$--terms that differ only in $m$.
The HEOM formulation developed by omitting such \emph{interference} ADOs
is termed as the sim-HEOM--$\alpha$; see Ref.~\onlinecite{Han18JCP}.
Therefore, the multi-level-and-multi-bath extension of Eq.~(11) 
is formally equivalent to the sim-HEOM--$\alpha$ method.

\section{Extension of Eq.~(11) of main text to spin-resolved cases}

For the single-impurity Anderson model studied in the main text,
the system involves explicitly the spin degree of freedom
(labeled by $s = \uparrow, \downarrow$).
The Grassmann-valued SEOM for $\trhos$ is
\begin{align}
 \dot{\tilde{\rho}}_{_{\rm S}} &= -i[\Hs, \trhos]
 + \sum_s \left[ B \big(\hc^\dag_s \,\eta_{1st} + \bareta_{2st}\, \hc_s \big) \,\trhos
 + B^{\ast} \trhos \big(\hc_s^\dag \,\eta_{3st} + \bareta_{4st} \,\hc_s \big) \right] \nl
  &\qquad + \sum_s B \,\big\{\hc_s^\dag\, \gm_{s}(t) - \gp_{s}(t)\, \hc_s, \,\trhos \big\},
  \label{eom-trhos-1}
\end{align}
where $\{g^\sigma_{s}(t)\}$ are given by
\begin{align}
  \gm_s(t) &=  \int_{t_0}^t
  \Big\{ [\Cp_s(t-\tau)]^\ast \, \eta_{4s\tau} - i \Cm_s(t-\tau)\,\eta_{2s\tau}  \Big\}\, d\tau,  \nl
  \gp_s(t) &=  \int_{t_0}^t
  \Big\{ [\Cm_s(t-\tau)]^\ast\, \bareta_{3s\tau} - i \Cp_s(t-\tau) \,\bareta_{1s\tau}  \Big\} \, d\tau. \label{gt-2}
\end{align}
Here, $\Cp_s(t-\tau) = \trb[\hF^\dag_s(t)\hF_s(\tau)\rhob^{\rm eq}]$
and $\Cm_s(t-\tau) = \trb[\hF_s(t)\hF^\dag_s(\tau)\rhob^{\rm eq}]$,
with $\hF_s(t) \equiv e^{i\int_{t_0}^t \!\Hb d\tau} \hF_s \, e^{-i\int_{t_0}^t \!\Hb d\tau}$
and $\hF_s = \sum_k t_{ks}\, \hat{d}_{ks}$.
By substituting the AGFs $\{\eta_{jst}, \bareta_{jst}\}$ with
\be
  \eta_{jst} \mapsto v_{jst} \Xd_{js}, \quad
  \bareta_{jst} \mapsto v_{jst} \Xu_{js},
\ee
\Eq{eom-trhos-1} is recast into the following numerically feasible form of
\be
  \dot{\tilde{\rho}}_{_{\rm S}} = -i [ \Hs, \trhos]
  + \sum_s \left(B\, \hc_s^\dag\, Y_{1s} \,\trhos
  + B\,Y_{2s} \,\hc_s\,\trhos
  + B^\ast \trhos\, \hc_s^\dag\,Y_{3s}
  + B^\ast \trhos\, Y_{4s}\,\hc_s \right),
  \label{eom-rhos-3}
\ee
where $\trhos = \sum_{l_{1\uparrow} \in S_{1\uparrow}}
\sum_{l_{1\downarrow} \in S_{1\downarrow}}
\sum_{l_{2\uparrow} \in S_{2\uparrow}}
\sum_{l_{2\downarrow} \in S_{2\downarrow}}
\trhos^{[l_{1\uparrow} l_{1\downarrow} l_{2\uparrow} l_{2\downarrow}]}$,
and 
\begin{align}
Y_{1s} &= v_{1st}\, \Xd_{1s} + \tgm_{st}\, \Xd_{2s},
   & Y_{2s} &= v_{2st}\, \Xu_{2s} - \tgp_{st}\, \Xu_{1s}, \nl
Y_{3s} &= v_{3st}\, \Xd_{1s} - i \tgm_{st}\,\Xd_{2s},
   & Y_{4s} &= v_{4st}\,\Xu_{2s} + i \tgp_{st}\, \Xu_{1s}.
\label{def-ys-1}
\end{align}
Here, the spin-resolved bath-induced stochastic fields $\{\tilde{g}^\sigma_{st}\}$ are
\begin{align}
  \tgm_{st} &= \int_{t_0}^t \left\{
  [C^+_s(t-\tau)]^\ast v_{4s\tau} - iC^-_s(t-\tau) v_{2s\tau} \right\} d\tau, \nl
  \tgp_{st} &=  \int_{t_0}^t \left\{
  [C^-_s(t-\tau)]^\ast v_{3s\tau} - iC^+_s(t-\tau) v_{1s\tau} \right\} d\tau.
  \label{def-tgt}
\end{align}
The physical reduced  system density matrix is finally obtained by
\be
  \rho = \la \trhos \ra = \mathcal{M}\left( \trhos^{[0000]}\right).
\ee
%

\bibliographystyle{aip}
\bibliography{bibrefs}